\documentclass[reprint,aip,apl,amsmath,amssymb]{revtex4-2}

\usepackage[dvipdfmx]{graphicx}
\usepackage{dcolumn}
\usepackage{bm}

\usepackage[utf8]{inputenc}
\usepackage[T1]{fontenc}
\usepackage{mathptmx}
\usepackage{textcomp}

\usepackage{color}

\begin{document}

\preprint{AIP/123-QED}

\title{Differential absorption ozone Lidar with 4H-SiC single-photon detectors}

\author{Xian-Song Zhao}

\affiliation{Hefei National Research Center for Physical Sciences at the Microscale and School of Physical Sciences, University of Science and Technology of China, Hefei 230026, China}
\affiliation{CAS Center for Excellence in Quantum Information and Quantum
Physics, University of Science and Technology of China, Hefei 230026, China}
\affiliation{Hefei National Laboratory, Hefei 230088, China}

\author{Chao Yu}
\altaffiliation{Authors to whom correspondence should be addressed: [Jun Zhang, zhangjun@ustc.edu.cn; Chao Yu, yuch@ustc.edu.cn]}
\affiliation{Hefei National Research Center for Physical Sciences at the
Microscale and School of Physical Sciences, University of Science and
Technology of China, Hefei 230026, China}
\affiliation{CAS Center for Excellence in Quantum Information and Quantum
Physics, University of Science and Technology of China, Hefei 230026, China}

\author{Chong Wang}
\affiliation{CAS Key Laboratory of Geospace Environment, University of Science
and Technology of China, Hefei 230026, China}

\author{Tianyi Li}
\affiliation{School of Electronic Science and Engineering, Nanjing University,
Nanjing 210023, China}

\author{Bo Liu}
\affiliation{Guoyao Quantum Lidar, Jinan 250013, China}

\author{Hai Lu}
\author{Rong Zhang}
\affiliation{Hefei National Laboratory, Hefei 230088, China}
\affiliation{School of Electronic Science and Engineering, Nanjing University,
Nanjing 210023, China}

\author{Xiankang Dou}
\affiliation{Hefei National Laboratory, Hefei 230088, China}
\affiliation{CAS Key Laboratory of Geospace Environment, University of Science
and Technology of China, Hefei 230026, China}

\author{Jun Zhang}
	\altaffiliation{Authors to whom correspondence should be addressed: [Jun Zhang, zhangjun@ustc.edu.cn; Chao Yu, yuch@ustc.edu.cn]}

\author{Jian-Wei Pan}

\affiliation{Hefei National Research Center for Physical Sciences at the
Microscale and School of Physical Sciences, University of Science and
Technology of China, Hefei 230026, China}
\affiliation{CAS Center for Excellence in Quantum Information and Quantum Physics, University of Science and Technology of China, Hefei 230026, China}
\affiliation{Hefei National Laboratory, Hefei 230088, China}

\date{\today}

\begin{abstract}
Differential absorption Lidar (DIAL) in the ultraviolet (UV) region is an effective approach for monitoring tropospheric ozone. 4H-SiC single-photon detectors (SPDs) are emergent devices for UV single-photon detection. Here, we demonstrate a 4H-SiC SPD-based ozone DIAL. We design
and fabricate the 4H-SiC single-photon avalanche diode with a beveled mesa
structure and optimized layer thickness. An active quenching circuit with a quenching time of
1.03 ns is developed to significantly mitigate the afterpulsing
effect while enhancing the maximum count rate. After characterization, the
SPD exhibits excellent performance with a photon detection efficiency of 16.6\%
at 266 nm, a dark count rate of 138 kcps, a maximum count rate of 13 Mcps, and
an afterpulse probability of 2.7\% at room temperature. Then, we apply two
4H-SiC SPDs in an ozone DIAL. The measured ozone concentrations at altitudes of 1-3.5 km agree well with the results of a commercial ozone DIAL. Our work provides an alternative solution for general UV Lidar applications.
\end{abstract}

\maketitle
Tropospheric ozone is an important atmospheric pollutant that is detrimental to biological health and contributes to the greenhouse effect\cite{JNC01,MRC09,DDF18}. Differential absorption Lidar (DIAL) in the ultraviolet (UV) band provides an effective approach for monitoring ozone concentration with the capability of long-term and high-resolution observation \cite{JTG14,MT16,RWR17}. Due to the strong scattering and absorption of UV lasers in the atmosphere, the backscattering signals of ozone DIAL attenuate rapidly with distance, leading to the requirement of highly sensitive and large-dynamic-range UV detectors. Currently, photomultiplier tubes (PMTs) are widely used in ozone DIALs. However, such devices suffer from intrinsic problems such as short lifetime, magnetic-sensitivity, and vacuum operation. In contrast, emergent 4H-SiC single-photon detectors (SPDs) have the advantages of quantum-limit sensitivity, small size, low cost, and ease-of-operation\cite{XXD07,LDH19,JCW23}, which make them promising candidates for practical UV Lidar applications.

4H-SiC SPDs comprise 4H-SiC single-photon avalanche diodes (SPADs) and readout circuits. Single-photon detection with 4H-SiC SPADs was first demonstrated in 2005\cite{XFP05}. Subsequently, various semiconductor structures and fabrication technologies have been proposed to improve the overall performance of 4H-SiC SPADs\cite{AGS05,XHD09,DFH14,FDH18}. However, 4H-SiC SPADs suffer from a relatively high afterpulse probability\cite{HHL20,YYY18}. Recently, we developed a dedicated passive quenching and active reset readout circuit and achieved an ultra-low afterpulse of 0.3\%, but, unfortunately, the maximum count rate (MCR) was less than 1 Mcps\cite{CTX23}. Low MCR and high afterpulse probability both lead to severe distortion in Lidar signals\cite{CMH17,CJH18,BCR24}. As a result, 4H-SiC SPDs have not yet been applied in high-precision Lidar applications.

\begin{figure*}[!t]\center
	\resizebox{17.5cm}{!}{\includegraphics{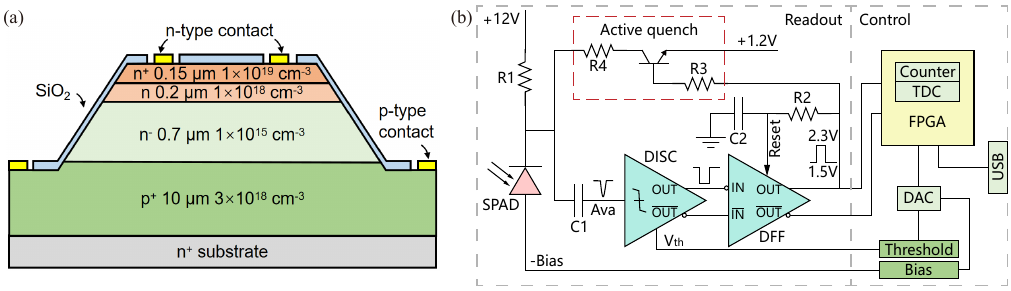}}
	\caption{Structure of the 4H-SiC SPAD and configuration of the 4H-SiC SPD:
	(a) Schematic diagram of the 4H-SiC SPAD structure with a beveled mesa;
	(b) Schematic circuit diagram of the 4H-SiC SPD.}
	\label{spd}
\end{figure*}

Afterpulse is generated by release carriers that are trapped by deep-level defects and impurities during previous avalanches. The afterpulse probability depends not only on the structural design and defect density of the device, but also on the readout circuit parameters. It can be roughly described by the following function\cite{JMH15}:
$P_{\mathrm{ap}}(t) \propto\left(C_{\mathrm{d}}+C_{\mathrm{p}}\right) \times
\int_0^\delta V_{\mathrm{ex}}(t) \mathrm{d} t \times e^{-\tau_{\mathrm{d}} /
\tau}$,
where $C_\mathrm{d}$ is the junction capacitance of the SPAD, $C_\mathrm{p}$ is the parasitic capacitance of the readout circuit, $\delta$ is the avalanche duration time, $\tau_{\mathrm{d}}$ is the hold-off time, and $\tau$ is the lifetime of trapped carriers. Therefore, the afterpulse probability can be mitigated by reducing the capacitance $C_\mathrm{d}$ and $C_\mathrm{p}$, shortening the avalanche duration time $\delta$, or controlling the carrier lifetime $\tau$.

Several readout techniques have been proven to effectively suppress afterpulse, such as high-speed gating circuits\cite{YWT20,YTW23,DSJ23}, negative feedback avalanche diodes (NFADs)\cite{BNT14,QC24}, and active quenching
circuits\cite{FAA13,JYY20}. In high speed gating circuits, the gating frequency always exceeds 1 GHz, therefore, the avalanche duration time $\delta$ can be suppressed to several hundred picoseconds. However, severe photon detection
efficiency (PDE) loss occurs when using such SPDs in free-running mode. NFAD
devices monolithically integrate a high-resistance thin-film resistor on the
surface of an SPAD, leading to a minimized parasitic capacitance
$C_\mathrm{p}$. Free-running SPDs based on NFAD devices have been widely used
in Lidar applications\cite{CMH17,CJH18,ZJX20,CJX21}. However, due to the large
integrated
resistor, the MCR of such SPDs is limited to less than 5 Mcps. In contrast, the
active quenching technique reduces the avalanche duration time $\delta$ by
using active feedback circuits. This technique can simultaneously achieve
free-running operation, low afterpulse probability, and high MCR, making it the
most appropriate approach for large-dynamic-range Lidar applications.

In this letter, we present a miniaturized free-running 4H-SiC SPD with
enhanced overall performance and demonstrate its application in ozone DIAL.
We design and fabricate the 4H-SiC SPAD with a beveled mesa structure and
optimized layers thickness. An active quenching readout circuit is designed to
simultaneously reduce afterpulse probability and enhance MCR. In the
experiments, our SPD exhibited a PDE of 16.6\%, a dark count rate (DCR) of 138 kcps, a MCR of 13 Mcps , and an afterpulse probability of 2.7\% at room temperature. A 5-hour continuous observation of
ozone concentration is conducted, and the results are compared with those of a
commercial ozone DIAL.

The cross-sectional schematic of the n-i-p 4H-SiC SPAD is shown in Figure~\ref{spd}(a). The initial epitaxial structure of the SPAD is grown on $4^{\circ}$ off-axis 4H-SiC n-type substrate, which comprises a heavily-doped $p^+$ buffer layer, a lightly doped $p^-$ absorption-multiplication layer, an n buffer layer, and an $n^+$ contact layer. The thickness of the absorption-multiplication layer is designed to be 0.7 \textmu m to ensure sufficient carrier acceleration distance, which guarantees a sufficient avalanche amplitude for discrimination. The fabrication process consists of mesa etching, ohmic contact
formation, and surface passivation\cite{CTX23}. A beveled mesa termination structure with a small slope angle of $\sim$$7^{\circ}$ is fabricated to suppress the peak electrical field around the mesa edge. The total thickness of the epitaxy layers is $\sim$11.05 \textmu m, and the diameter of the active area is $\sim$230 \textmu m. The breakdown voltage of the SPAD is $\sim$230 V, and the maximum avalanche gain could exceed $10^5$. The peak spectral responsivity of the SPAD is located at 280 nm, featuring a peak quantum efficiency of $\sim$52\%.

\begin{figure}[t!]
	\centering\includegraphics[width=8.5cm]{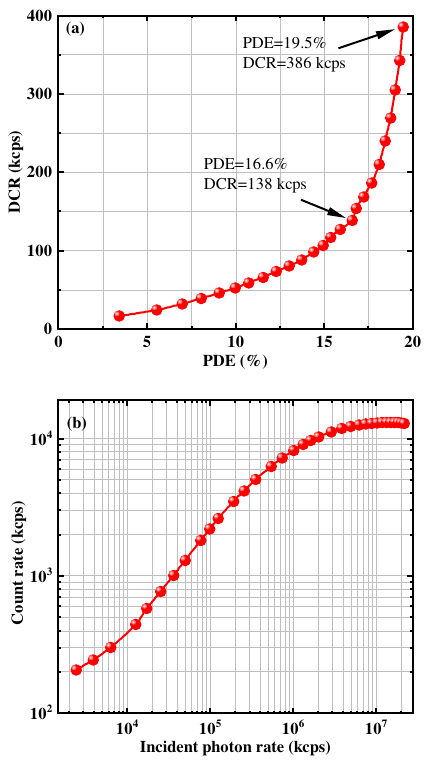}
	\caption{Performance calibration results of the 4H-SiC SPD:
		(a) Dark count rate versus photon detection efficiency at 266 nm;
		(b) Measured count rate versus incident photon number.}
	\label{calibration}
\end{figure}

Using the 4H-SiC SPAD, we further develop the free-running 4H-SiC SPD. As
shown in Figure~\ref{spd}(b), the 4H-SiC SPD comprises a readout circuit and a
control circuit. In the readout circuit, the SPAD cathode is connected to a
fixed 12 V voltage via a 500 $\Omega$ resistor (R1), and the SPAD anode is
connected to an adjustable negative bias voltage. The avalanche signal is
extracted from the cathode of the SPAD via a 10 pF capacitor (C1). Then, the signal
is discriminated to standard low voltage positive emitter coupled logic
(LVPECL), and the pulse width is regulated to $\sim$60 ns via a D-type
flip-flop (DFF). The regulated signal is connected to a field
programmable gate array (FPGA) as the detection output signal, whereas
the positive part of the differential detection signal is used to turn on a
high-speed bipolar junction transistor (BJT) to immediately pull the SPAD cathode
voltage to 1.2 V. The quenching time, including the delay time
of the discriminator, DFF, and active quenching circuit, is measured as
1.03($\pm$0.05) ns.
After a period of hold-off time, the output of the DFF is reset to logical 0, which
turns off the BJT and rearms the SPAD for subsequent detection.

In the control circuit, an FPGA is used to process the photon detection signal, communicate with a personal computer (PC), and set the SPD parameters. For signal processing, a counter module is developed in the FPGA to monitor the count rate of the SPD. In addition, a time-to-digital converter (TDC) module with an accuracy of 10 ns is also developed to measure the arrival time of photon detection events. The SPD communicates with the PC via a universal serial bus (USB) interface. The count rate and TDC data are uploaded to a PC in real time, and the SPD parameters, including bias voltage and threshold voltage, are downloaded from the PC and set via a digital-to-analog converter (DAC). The size of the SPD is 112$\times$88$\times$50 $mm^3$, and its weight is 0.35 kg.

Following the method\cite{CTX23}, we calibrate the performance of the 4H-SiC SPD in terms of PDE, DCR and afterpulse probability. A pulsed laser emits a collimated beam with a pulse width of 100 ps and a repetition frequency of 50 kHz at 266 nm. Then, the beam is divided by a beam splitter, with one channel incident to a power meter for monitoring, and the other channel attenuated to one photon per pulse. The attenuated beam is then deflected by a scanning galvo and focused on the active area of the 4H-SiC SPD via an aspheric lens. Finally, the interval time between the laser pulse and detection events is measured by a TDC. As the delay from laser pulse to photon detection event is constant, the photon detection signal, dark count, and afterpulse events can be easily distinguished. Thus, the PDE, DCR, and afterpulse probability can be precisely estimated according to the TDC data.

Figure~\ref{calibration}(a) shows the measured DCR as a function of the PDE at 266
nm and a room temperature of 20 $^{\circ}$C. The DCR rapidly increases with the PDE.
The maximum PDE of the 4H-SiC SPD is approximately 19.5\%, in which case the DCR
reaches 386 kcps. In the experiments, to obtain an optimized overall
performance, 16.6\% PDE and 138 kcps DCR are selected as the working point. The
results of detection counts versus incident photon number are shown in
Figure~\ref{calibration}(b). As MCR is independent of PDE, the results are
measured at a PDE of $\sim$3\% to avoid damaging the SPD under strong incident light. The results show that when the incident photon rate is greater than 10 Gcps, the MCR reaches 13 Mcps.

\begin{figure}[!t]
	\centering\includegraphics[width=8.5cm]{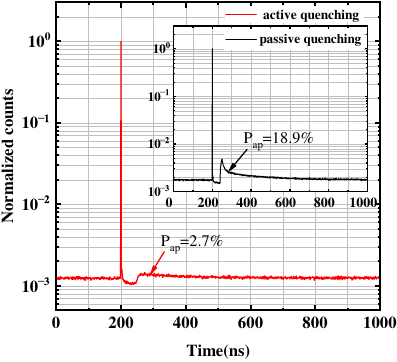}
	\caption{Normalized TDC data measured at a PDE of 16.6\%.
    With 1.03 ns active quenching, the afterpulse probability is 2.7\%. The inset shows the results when the active quenching circuit is disconnected and passive quenching is used, the afterpulse probability increases to 18.9\%.}
	\label{afterpulse}
\end{figure}

Figure~\ref{afterpulse} shows the typical normalized TDC histogram at the 16.6\% PDE working
point. The first peak corresponds to photon counts of the laser
pulse, and the second small peak after a 60 ns hold-off time corresponds to the
afterpulsing counts. With the active quenching circuits connected, the total
afterpulse probability is 2.7\%. In comparison, when the active
quenching circuit is disconnected and only passive quenching is used, the total afterpulse
probability increases to 18.9\%, as shown in the inset of Figure~\ref{afterpulse}.
The results prove that the afterpulse probability decreases with the reduce of avalanche time reduction, giving a 7 times decrease when employing 1.03 ns active quenching circuit compared to passive quenching. The
overall performance of the SPD (i.e., a PDE of 16.6\%, a DCR of 138 cps, an afterpulse probability of 2.7\%, and an MCR of 13 Mcps) meets the requirements of most Lidar applications.

\begin{figure}[!t]
	\centering\includegraphics[width=8cm]{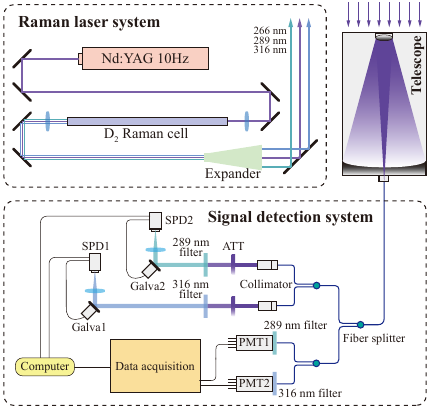}
	\caption{Experimental setup of ozone DIAL system using 4H-SiC SPDs and
	PMTs. The Raman laser system generates 289 nm and 316 nm laser beams and
	sends them to the atmosphere vertically. A telescope collects the backscattering
	signal to a multi-mode fiber. The signal is equally divided into two
	channels, and then detected by the 4H-SiC SPD system and PMT system,
	respectively.}
	\label{setup}
\end{figure}

We then demonstrate the application of 4H-SiC SPDs based on a commercial ozone DIAL system, which features a temporal resolution of 15 min and a spatial resolution of 60 m. The experimental setup is shown in Figure~\ref{setup}, which
includes a Raman laser system, telescope, and signal detection system. In the
Raman laser system, a Nd:YAG laser at 266 nm emits a laser beam with an energy
of 100 mJ per pulse and a frequency of 10 Hz. Then, the wavelength of the laser
is modulated to 289 nm and 316 nm through a deuterium Raman cell. According to the spectral response of the 4H-SiC SPAD, when the SPD exhibits an efficiency of 16.6\% at 266 nm, the PDE at 289 nm and 316 nm are estimated to be $\sim$ 18\% and 12\%, respectively.
The three beams are expanded to a diameter of 75 mm and a
divergence angle of 0.5 mrad via a beam expander, and finally sent to the
atmosphere vertically.

The backscattering signal from the atmosphere is collected into a multi-mode fiber, with the receiving telescope's field of view (FOV) covers the transmitting FOV at altitudes above 1 km. In the signal detection system, the backscattering signal is divided
equally into two channels: one channel is detected by our 4H-SiC SPDs, and the
other channel is connected to the original PMT system for comparison. In the
SPD channel, the signal is further divided by a 1:1 beam splitter, and then
sent to the free-space via a pair of collimators. The signal power is optimized
by adjustable attenuators to avoid SPD saturation. The lasers then pass through a 289 nm bandpass filter with a bandwidth of 1.24 nm and a transmittance of 0.59, and a 316 nm bandpass filter with a bandwidth of 1.02 nm and a transmittance of 0.68, respectively.
Finally, two galvanometers and aspherical lenses are used to focus the beams on
the active area of 4H-SiC SPDs. The flying time of the backscattering signals
is measured by the integrated TDCs and sent to the PC every five minutes. In
the PMT channel, the signal is also divided and filtered to 289 nm or 316 nm
parts, and detected by two PMTs. The data acquisition system measures the
time-correlated output current intensity and sends the data to the PC.

\begin{figure*}[!t]\center
	\resizebox{17.5cm}{!}{\includegraphics{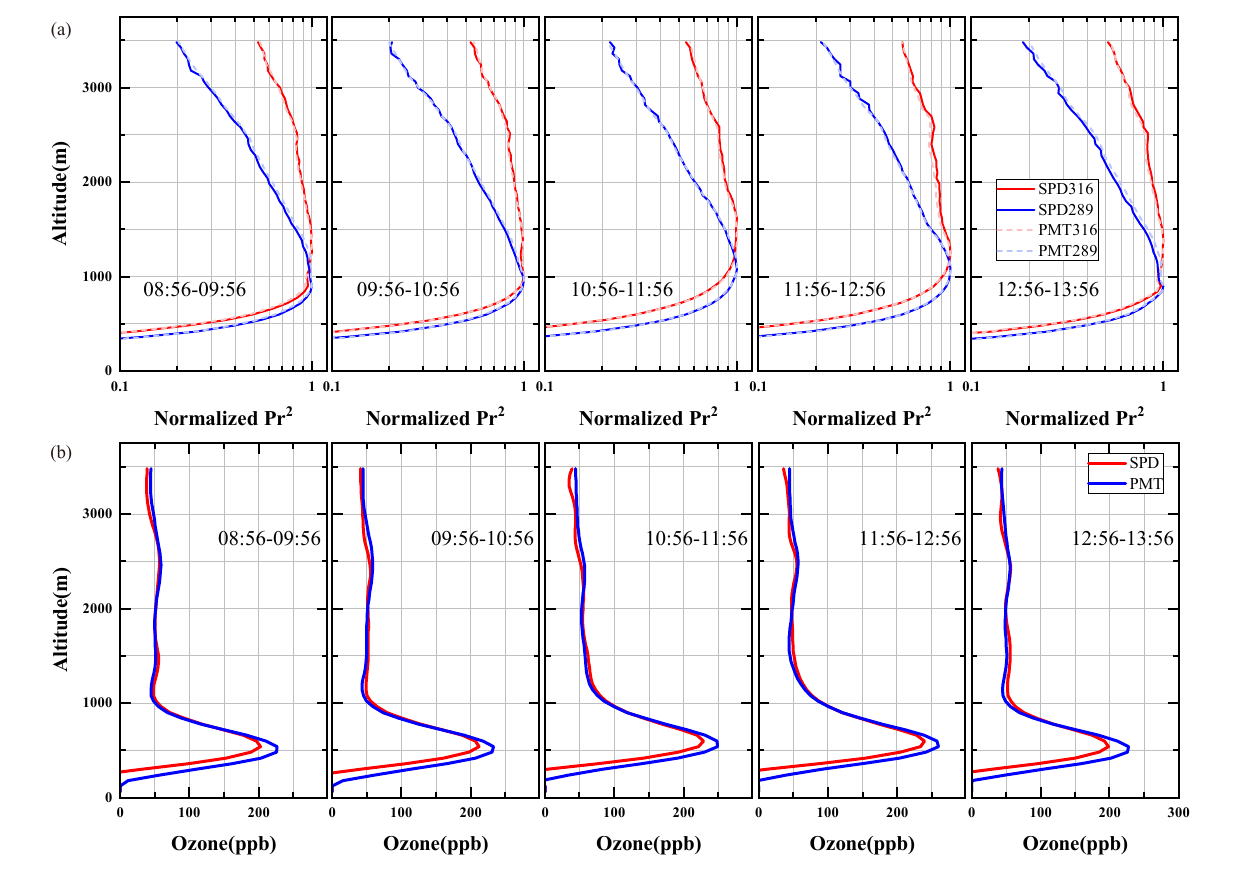}}
	\caption{Continuous observation of ozone concentration measured by SPDs and
	PMTs simultaneously.
	(a) Normalized $Pr^2$ versus altitude. For the SPD data, hold-off time correction, afterpulse correction, and DCR correction have been performed~\cite{CMH17}. For the PMT data, the dark current correction has been performed.
	(b) Ozone retrieval results.
	The data was measured on September 14, 2023, in Jinan, China.}
	\label{ozone}
\end{figure*}

In ozone DIAL, according to the Lidar equation, assuming that the emitting power of the laser pulse is $P_L(\lambda)$, the power of the received backscattering signal at a distance $R$ can be calculated as:

\begin{equation}
\begin{aligned}
P(\lambda, R)= & C(R) P_{\mathrm{L}}(\lambda) \frac{A}{R^2} \Delta R
\beta(\lambda, R) \\
& \exp \left(-2 \int_0^R[\alpha(\lambda, R)+\sigma N(R)] \mathrm{d} r\right)
\end{aligned}
\label{lidar}
\end{equation}
where $C(R)$ is the overlap factor between the laser beam and the receiver field of view, $A$ is the effective receiving area, $\Delta R$ is the range resolution of the Lidar, $\beta(\lambda, R)$ is the backscattering coefficient of the atmosphere, $\alpha(\lambda, R)$ is the extinction coefficient of the atmosphere, $\sigma$ represents the absorption cross section of ozone, which equals $1.59 \times 10^{-18} cm^2$ and $4.38 \times 10^{-20} cm^2$ at 289 nm and 316 nm, respectively. $N(R)$ represents the ozone concentration.

In the experiments, we assume that the atmosphere backscattering coefficient $\beta(\lambda, R)$ and extinction coefficient $\alpha(\lambda, R)$ are almost the same at wavelengths of 289 nm and 316 nm, while the effect of temperature on atmospheric parameters is not take into account. Then, the ozone concentration can be derived from $P(\lambda_{289},R)$ and $P(\lambda_{316},R)$ as:

\begin{equation}
N(R)=\frac{1}{2 \Delta \sigma} \frac{d}{d R} \ln
\frac{P(\lambda_{316},R)}{P(\lambda_{289},R)}
\label{density}
\end{equation}
where $\Delta\sigma=\sigma_{289}-\sigma_{316}$ is the differential absorption cross section.

For the SPD data, we apply hold-off time correction, afterpulse correction, and DCR correction following the method~\cite{CMH17}. For the PMT data, the dark current correction is performed. As the laser repetition frequency is relatively low in this demonstration experiment, the shot noise of the original photon counts is non-negligible, which introduces a severe error in the ozone concentration calculation. Therefore, we use a Gauss weighted moving average filter to smooth the data vector. In the algorithm, the photon count at a certain altitude is corrected as the average of the nearest 15 points with a Gaussian weight. After smoothing, the ozone concentrations can be precisely calculated by Eq.~\ref{density}.

Figure~\ref{ozone} shows 5 h of continuous observations from 8:56 to 13:56
on September 14, 2023. Figure~\ref{ozone} (a) shows the normalized $Pr^2$ measured by the 4H-SiC SPD (solid line) and the PMT (dotted line), both of which have been processed by the correction algorithm. Since ozone absorbs more strongly at 289 nm, the echo signal at this wavelength decreases more rapidly with height compared to the signal at 316 nm. For a given wavelength, the echo signal measured by the SPD and the PMT is nearly identical.
The spatial resolution is set to 60 m, and the data are cumulated in 1 h
increment. The
overlap factor of the telescope system reaches 100\% at altitudes above 1 km.
Figure~\ref{ozone} (b) shows the retrieved ozone concentration.
During observation, the ozone concentration fluctuates around 50 ppb within
altitudes ranging from 1 km to 3.5 km. The derived concentrations and varying
trends of the two systems agree very well.
The results prove that the present
4H-SiC SPDs can be used in high-precision Lidar applications. In the future, by
applying a high-repetition-frequency laser source and operating the 4H-SiC SPD
at lower temperatures, the resolution and accuracy of ozone DIAL can be
significantly improved.

In summary, we present a high-performance 4H-SiC SPD with a high MCR and low afterpulse probability, and apply 4H-SiC SPDs in an ozone DIAL system. This is achieved by designing and fabricating a 4H-SiC SPAD with an optimized semiconductor structure, while developing an active quenching circuit with a quenching time of 1.03 ns . The 4H-SiC SPD exhibits excellent performance with a PDE of 16.6\% at 266 nm, a DCR of 138 kcps, an MCR of 13 Mcps, and an afterpulse probability of 2.7\% at room temperature. We apply the 4H-SiC SPD in a commercially available ozone DIAL system, and compare the ozone concentrations measured by our 4H-SiC SPD and the original PMT system simultaneously. The results agree well at altitudes of 1-3.5 km. Considering the advantages of the small-size, low-cost, and high-stability of the 4H-SiC SPD, our work provides an alternative solution for general UV Lidar applications.

This work was supported by the Innovation Program for Quantum Science and Technology under Grant 2021ZD0300804, 2021ZD0303400, and the National Natural Science Foundation of China (Grant No. 62175227).


~\\
\textbf{AUTHOR DECLARATIONS}

\noindent{\textbf{Conflict of Interest}}
The authors have no conflicts to disclose.

~\\
\textbf{DATA AVAILABILITY}

The data that support the findings of this study are available from the corresponding author upon reasonable request.



\section*{REFERENCES}
\begin{thebibliography}{0}
	
	\bibitem{JNC01}
	J. Staehelin, N. R. P. Harris, C. Appenzeller, and J. Eberhard, Rev. 	
	Geophys.
	\textbf{39}, 231-290 (2001).
	
	
	\bibitem{MRC09}
	M. Jerrett, R. T. Burnett, C. A. Pope III, K. Ito, G. Thurston, D. Krewski,
	Y. Shi, E. Calle, and M. Thun, N. Engl. J. Med.
	\textbf{360}, 1085-1095 (2009).
	
	\bibitem{DDF18}
	D. Nuvolone, D. Petri, and F. Voller, Environ. Sci. Pollut. Res.
	\textbf{25}, 8074-8088 (2018).
	
%
%
%
\bibitem{JTG14}
	J. T. Sullivan, T. J. McGee, G. K. Sumnicht, L. W. Twigg, and R. M. Hoff,
	Atmos. Meas. Tech.
	\textbf{7}, 3529–3548 (2014).
	
	\bibitem{MT16}
	M. J. Granados-Muñoz, and T. Leblanc, Atmos. Chem. Phys.
	\textbf{16}, 9299–9319 (2016).

    \bibitem{RWR17}
	R. De Young, W. Carrion, R. Ganoe, D. Pliutau, G. Gronoff, T. Berkoff, and
	S. Kuang, Appl. Opt.
	\textbf{56}, 721-730 (2017).

\bibitem{XXD07}
	X. Bai, X. Guo, D. C. Mcintosh, H.-D. Liu and J. C. Campbell,
	IEEE J. Quantum Electron.
	\textbf{43}, 1159-1162 (2007).
	
	
	\bibitem{LDH19}
	L. Su, D. Zhou, H. Lu, R. Zhang, and Y. Zheng, J. Semicond.
	\textbf{40}, 121802 (2019).

    \bibitem{JCW23}
    J.-T. Ye, C. Yu, W. Li, Z.-P. Li, H. Lu, R. Zhang, J. Zhang, F. Xu, and J.-W. Pan, Appl. Phys. Lett. \textbf{123}, 024005 (2023).

\bibitem{XFP05}
	X. Xin, F. Yan, P. Alexandrove, X. Sun, C.M. Stahle, J. Hu, M. Matsumura,
	X. Li, M. Weiner, and H.J. Zhao, Electron. Lett.
	\textbf{41}, 212-214 (2005).

	\bibitem{AGS05}
	A. L. Beck, G. Karve, S. Wang, J. Ming, X. Guo, and J. C. Campbell, IEEE
	Photonics Technol. Lett.
	\textbf{17}, 1507-1509 (2005).
	
	\bibitem{XHD09}
	X. Bai, H.-D. Liu, D. C. McIntosh, and J. C. Campbell, IEEE J. Quantum
	Electron.
	\textbf{45}, 300-303 (2009).
	
	\bibitem{DFH14}
	D. Zhou, F. Liu, H. Lu, and D. Chen, IEEE Photon. Technol. Lett.
	\textbf{26}, 1136-1138 (2014).
	
	\bibitem{FDH18}
	X. Cai, L. Li, H. Lu, D. Zhou, W. Xu, D. Chen, F. Ren, R. Zhang, Y. Zheng,
	and G. Li, IEEE Photon. Technol. Lett.
	\textbf{30}, 805-808 (2018).

\bibitem{YYY18}
Y. Wang, Y. Lv, Y. Wang, Q. Zhang, S. Yang, D. Zhou, H. Lu, E. Wu, and G. Wu, IEEE Journal of Selected Topics in Quantum Electronics \textbf{24}, 1-5 (2018).

\bibitem{HHL20}
H. Dong, H. Zhang , L. Su, D. Zhou, W. Xu , F. Ren, D. Chen, R. Zhang, Y. Zheng, and H. Lu, IEEE Photonics Technology Letters \textbf{32}, 706-709 (2020).
	
	\bibitem{CTX23}
	C. Yu, T. Li, X.-S. Zhao, H. Lu, R. Zhang, F. Xu, J. Zhang, and J.-W Pan,
	Rev. Sci. Instrum.
	\textbf{94}, 033101 (2023).

\bibitem{CMH17}
	C. Yu, M. Shangguan, H. Xia, J. Zhang, X. Dou, and J. -W. Pan, Opt. Express
	\textbf{25}, 14611-14620 (2017).

\bibitem{CJH18}
C. Yu, J. Qiu, H. Xia, X. Dou, J. Zhang, and J.-W. Pan,  Rev. Sci. Instrum. \textbf{89}, 103106 (2018).

\bibitem{BCR24} 
B. Yang, C. Wang, R. Zhao, X. Xue, T. Chen, and X. Dou, Opt. Express \textbf{23}, 11992-12003 (2024).

\bibitem{JMH15}
	J. Zhang, M. A Itzler, H. Zbinden, and J. -W. Pan, Light Sci. Appl.
	\textbf{4}, e286 (2015).


\bibitem{YWT20}
Y. Fang, W. Chen, T.-H. Ao, C. Liu, L. Wang, X.-J. Gao, J. Zhang, and J.-W. Pan, Rev. Sci. Instrum. \textbf{91}, 083102 (2020).

\bibitem{YTW23}
Y. Fan, T. Shi, W. Ji, L. Zhou, Y. Ji, and Z. Yuan, Opt. Express \textbf{31}, 7515-7522 (2023).

\bibitem{DSJ23}
D.-Y. He, S. Wang, J.-L. Chen, W. Chen, Z.-Q. Yin, G.-J. Fan-Yuan, Z. Zhou, G.-C. Guo, and Z.-F. Han, Adv. Devices Instrum. \textbf{4}, 0020 (2023).
	
    \bibitem{BNT14}
    B. Korzh, N. Walenta, T. Lunghi, N. Gisin, and H. Zbinden, Appl. Phys. Lett. \textbf{104}, 081108 (2014).

	\bibitem{QC24}
    Q. Xu, C. Yu, W. Chen, J. Zhao, D. Cui, J. Zhang, and J.-W. Pan, IEEE J. Sel. Top. Quant. \textbf{30}, 6400107 (2024).
	
	
	\bibitem{FAA13}
	F. Acerbi, A. D. Frera, A. Tosi, and F. Zappa, IEEE J. Quantum Electron.
	\textbf{49}, 563-569 (2013).

    \bibitem{JYY20}
	J. Liu, Y. Xu, Y. Li, Y. Gu, Z. Liu and X. Zhao, J. Mod. Optic. \textbf{67}, 1184-1189 (2020).

	\bibitem{ZJX20}
	Z.-P. Li, J.-T. Ye, X. Huang, P.-Y. Jiang, Y. Cao, Y. Hong, C. Yu, J.
	Zhang, Q. Zhang, C.-Z. Peng, F. Xu, and J.-W. Pan, Optica
	 \textbf{8}, 344-349 (2021).
	
	\bibitem{CJX21}
	C. Wu, J. Liu, X. Huang, Z.-P. Li, C. Yu, J.-T. Ye, J. Zhang, Q. Zhang, X.
	Dou, V.K. Goyal, F. Xu, J.-W. Pan, Proc. Natl. Acad. Sci. U.S.A.
	\textbf{118}, e2024468118 (2021).


\end{thebibliography}

\end{document}